\documentclass[aps,twocolumn,showpacs, prl,10 pt]{revtex4-1}

\usepackage{lmodern}
\usepackage[T1]{fontenc}
\usepackage[utf8]{inputenc}
\usepackage{braket}
\usepackage{leftidx}
\usepackage{color}         
\usepackage{subfigure}  
\usepackage{dcolumn}    
\usepackage{textcomp}
\usepackage{lipsum}
\usepackage{threeparttable}
\usepackage{multirow}
\usepackage{siunitx}
\usepackage{tabularx}
\usepackage{graphicx,amsmath,amssymb,bm}
\usepackage{hyperref,verbatim}
\usepackage{bbold}
\hypersetup{%
  breaklinks = {true},
  citecolor = {blue},
  colorlinks = {true},
  linkcolor = {blue},
  pdfauthor = {\textcopyright\ Ali Hallal},
  pdfcreator = {\LaTeX\ and \flqq hyperref\frqq},
  pdffitwindow = {true},
  pdfmenubar = {true},
  pdfpagelayout = {SinglePage},
  pdfstartview = {Fit},
  pdftoolbar = {true},
  plainpages = {false},
}

\begin{document}

\title{ Graphene-based spinmechatronic valve}

\author{Ali Hallal} 
\affiliation{Univ. Grenoble Alpes, CEA, CNRS, Spintec, 38000 Grenoble, France}%
\email{ali.hallal@hotmail.fr, ali.hallal@cea.fr}

\begin{abstract}
Interlayer twist between van der Waals graphene crystals led to the discovery of superconducting and insulating states near the magic angle. In this work, we exploit 
this mechanical degree of freedom by twisting the graphene middle layer in a trilayer graphene spacer between two metallic lead (Magnetic and nonmagnetic).
A large difference in conductance is found depending on the angle of twist between the middle layer graphene and the ones at the interface this difference ,called 
twisting resistance, reach more than 1000\% in the non-magnetic Cu case. For the magnetic Ni case, the  magneto-resistance decreases and the difference in conductance 
between twisted and not twisted depends strongly on the  relative magnetization configuration. For the parallel configuration, the twisting resistance is about -40\%, 
while for the anti-parallel configuration it can reach up to 130\% . Furthermore, we show that the twisting resistance can be enhanced by inserting a thin Cu layer at 
the interface of Ni/graphene where it reaches a value of 200\% and 1600\% for parallel and antiparallel configurations, respectively. These finding could pave the way toward
 the integration of 2D materials on novel spinmechatronics based devices.
\end{abstract}


\maketitle


Graphene incorporation in magnetic tunnel junctions (MTJs)~\cite{Cobas2012,Singh2014,Cobas2016,HU2019} has attract a lot of attention recently due to 
its protective nature against the oxidation of the ferromagnetic layer~\cite{Piquemal-Banci2018}. In addition graphene multilayers are found to be a perfect spin 
filter in vertical magnetic tunnel junctions~\cite{Karpan2007,Saha2012,QIU2019622}. Furthermore, and despite its weak spin-orbit coupling (SOC), graphene coating 
of Co is found to induce a large perpendicular magnetic anisotropy~\cite{Vo_Van_2010, Rougemaille2010, Coraux2012, Yang2016} and a significant Dzyaloshinskii-Moriya interaction due to the Rashba effect~\cite{Yang2018,Ajejas2018,Chaurasiya2019}. 
In addition, graphene-based magnetic multilayers show a strong perpendicular anti-ferromagnetic exchange coupling that could be used as synthetic anti-ferromagnetic~\cite{Barla2016,Gargiani2017}.  Those 
findings have promoted graphene, and subsequently other two dimensional (2D) materials, as potential candidates to replace oxide barrier in conventional MTJs.

The weak coupling between different layers of 2D materials allows to manipulate the relative twisting angle (RTA) between two or more layers of graphene leading to the formation of moir\'e superlattices. 
The electronic properties of such herterostructure strongly depend on the RTA between two or more layers. This led recently to the discovery of superconductivity~\cite{Cao2018Natur,Kerelsky2019,Chen2019,Isobe2018}, 
insulating state~\cite{Cao2018is,Yonglong2019,Kang2018,Koshino2018}, and even orbital ferromagnetism~\cite{Sharpeeaaw3780, Liu2019} near the "magic angle" in twisted bilayer graphene (TBLG). Furthermore, recent experiment show the possibility 
to control in-situ~\cite{Ribeiro-Palau690} the RTA and thus giving rise to a mechanical degree of freedom led to an emerging field called twistronics. The transport properties of TBLG~\cite{Peeters2018,Chung2018,Hwang2019} 
and bilayer graphene nanoribbons~\cite{Brandimarte2017} has been lately investigated in lateral geometry show a strong effect of the RTA on the electric properties of the system. A scientific investigation of transport 
properties in vertical geometry is still lacking. Incorporating graphene in MTJ and varying the RTA will strongly affects the electronic properties of MTJs; a procuration which is not possible using conventional oxide structures.

\begin{figure*}[htp]
\centering
\includegraphics[clip = true, trim=0cm 0.0cm 0.0cm 0.0cm, width=1.0\textwidth]{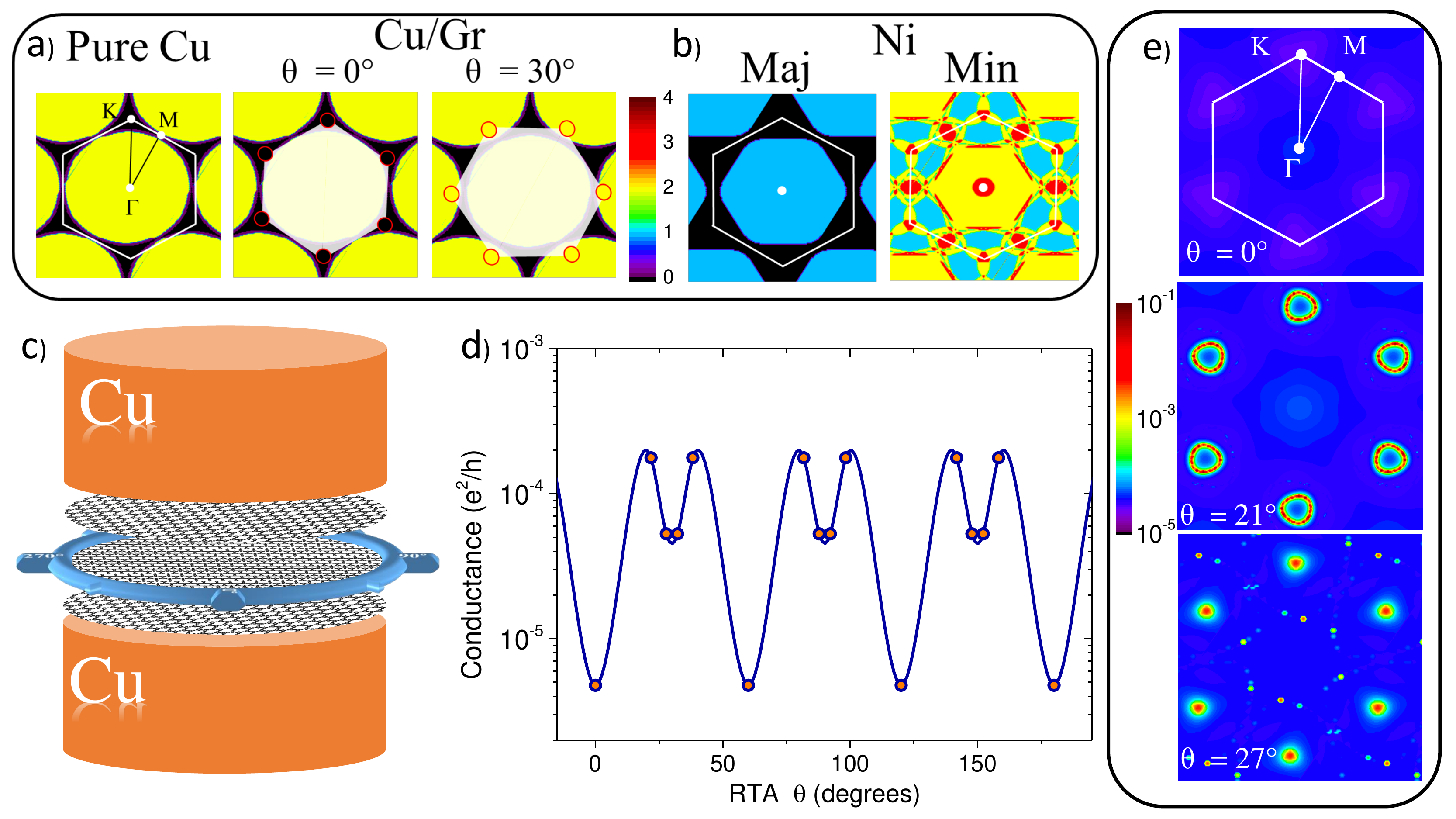}
\caption{ (a) (left panel) Transmission map across the Brillouin zone of Pure Cu, middle and right panel show a schematic of graphene Fermi surface in top of Cu transmission map with a 
RTA angle of 0$^{\circ }$ and 30$^{\circ }$, respectively. (b) Majority and minority transmission map across the Brillouin zone of Ni. (c) Schematic of the proposed device composed of Cu/TTLG/Cu 
junction where the middle layer of graphene could be twisted relative to the ones at the interfaces. (d) Quantum conductance at room temperature of Cu/TTLG/Cu as function of the RTA. (e) top middle 
and bottom show transmission map across Brillouin zone at Fermi energy for RTA of 0$^{\circ }$, 21.787$^{\circ }$ and 27.8$^{\circ }$, respectively. The Brillouin zone with high symmetry points are shown in white.}
\label{1}
\end{figure*}

In this work, using first principle calculations, we demonstrate a significant difference in resistance between non-twisted and twisted tri-layer graphene (TTLG) spacer, we call as twisting-resistance (TwR), 
between two metallic leads (magnetic or non-magnetic). In case of non-magnetic Cu substrate, the TwR could reach more than 1000\%  which is attributed to the absence of conducting state in large region around 
the K point in reciprocal space. Once the middle layer of graphene is twisted, it start to feel the populated states present round the M point and thus a large current go through the graphene layer. For the 
magnetic Ni case, the current are mostly dominant by the minority electrons for parallel configuration as explained and shown earlier by V. M. Karpan et. al.~\cite{Karpan2007}, leading to perfect spin filter. 
Rotating the middle layer lead to a reduction of the spin filter effect due to the increase of majority conductance.

First, we start by examining the transmission map across the Brillouin zone at Fermi level for pure Cu film as shown in Fig.~\ref{1} (a) (left panel), the current passes homogeneously throughout the reciprocal 
space except at the edges where there is no state available. In the other hand, graphene or graphite the only state close to Fermi energy are found near the high symmetry K point in reciprocal space. So if we 
put graphene or graphite in top of Cu the electrons will have to overcome a large barrier to tunnel Fig.~\ref{1} (a) (middle panel). However, if graphene layer is twisted by an angle of 30$^{\circ }$ the Dirac 
cone of graphene are now in top or close to the region where the Cu have high transmission probability  and thus the electrons will go through easily. For a ferromagnetic case ,such as Ni or Co, the  only states 
are present around the K point are the minority spin character while for majority spin case the transmission map is similar to that of Cu and this fact is at the origin of large spin filtering effect in 
Fm/graphite/FM (FM = Ni or Co) [c.f. Fig.~\ref{1} (b) ]. Similarly, If graphene layer in top of Ni is twisted we expect the Majority current will enhance and that of the minority will decreases. 

\begin{figure*}[htp]
\centering
\includegraphics[clip = true, trim=0cm 0.0cm 0cm 0.0cm, width=1.0\textwidth]{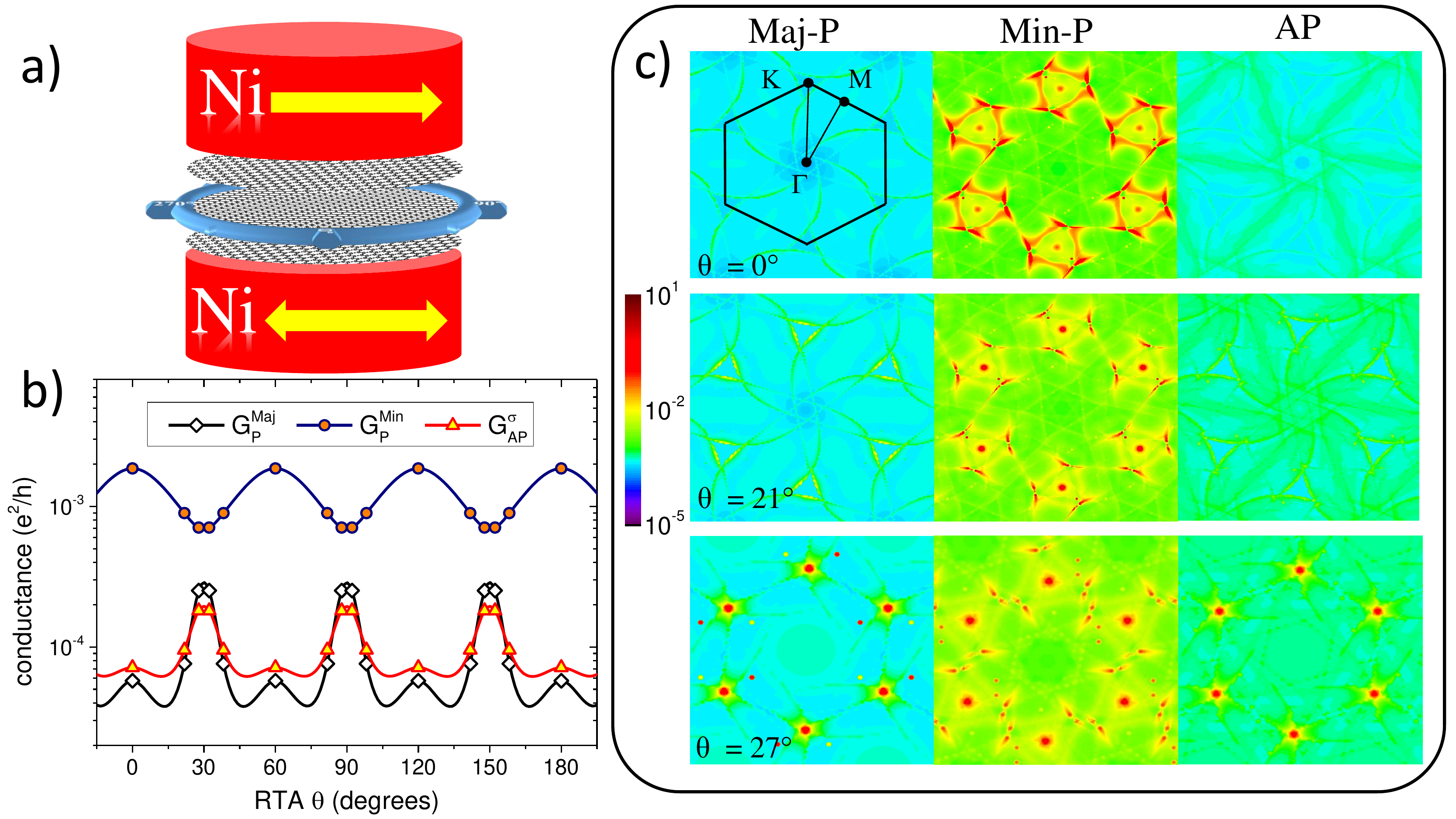}
\caption{ (a) Device schematic composed of Ni/TTLG/Ni where the middle layer graphene is rotated relative to the ones at the interface. (b)  Majority, minority conductance for parallel and anti-parallel 
configurations of Ni/TTLG/Ni MTJs at room temperature as function of the RTA. (c)  Transmission map across  Brillouin zone at Fermi energy for majority parallel (left), minority parallel (middle) and 
anti-parallel (right) for RTA of 0$^{\circ}$ (top), 21.787$^{\circ}$ (middle) and 27.8$^{\circ}$ (bottom).   }
\label{2}
\end{figure*}

To verify our hypothesis and to investigate the effect of RTA on the electronic transport properties in vertical  graphene-based junctions, we setup a structure of Cu/TTLG/Cu, where only 
the middle layer graphene is rotated relative to the graphene layers at the interface as shown in Fig.~\ref{1} (c). There are two types of twisted layer of graphene in the literature the 
commensurate and non-commensurate~\cite{Yao2018,Morell2013} in this paper we focus on the commensurate twisted layer graphene. Due to computational limitation we can only study two relative 
rotation angles (RTA) of commensurate twisted graphene that have the smallest possible unit cell. The two RTA we study are 21.787$^{\circ}$  and 27.8$^{\circ}$ that correspond to a unit cell 
of $\sqrt{7}\times\sqrt{7}$ and $\sqrt{13}\times\sqrt{13}$, respectively. Due to the six fold symmetry of graphene, the 21.787$^{\circ}$ and 27.8$^{\circ}$ is similar to that of 38.213$^{\circ}$ 
and 32.2$^{\circ}$.  The in-plane lattice constant and the interlayer distance between graphene and substrate are taken from ref.~\cite{Karpan2007} The scattering region is composed of trilayer 
of graphene sandwiched between 6 layers of Cu or Ni and attached to 6 layers of Cu or Ni lead.

Our first-principles calculations were performed using  SIESTA package~\cite{Artacho1999,Soler2002} and the exchange correlation energy calculated within the generalized gradient approximation of 
Perdew-Burke-Ernzerhof (PBE)~\cite{Perdew1996} and valence states were expanded by numerical atomic orbital basis sets with Single-zeta (SZ) functions. Atomic cores were described by non-local norm 
conserving Troullier-Martins pseudopotentials. The cutoff energy is set to 300 Ry and found to be more than sufficient in our case. For RTA of 21.787$^{\circ}$ (32.2$^{\circ}$) Monkhorst-Packgrids 
of 7$\times$7$\times$1 (3$\times$3$\times$1) and 7$\times$7$\times$20 (3$\times$3$\times$20) k-points were used for the Brillouin zone integration of the device and electrodes, respectively.  Nonequilibrium 
Green function (NEGF) transport calculation were performed using TranSIESTA ~\cite{Brandbyge2002,PAPIOR2017} to study spin-dependent transport. For transmission calculation a 100$\times$100$\times$1 and 
50$\times$50$\times$1 is used for RTA of 21.787$^{\circ}$ and 32.2$^{\circ}$ using the TBTrans code, respectively.

In general the twisting magneto-resistance can be defined as the difference between two conducting state with/without different twisting angle and/or different magnetization orientation, as following:

\begin{equation*}
\left(\frac{\Delta G}{G}\right)_{\alpha,\alpha '} (\theta, \theta ') = \frac{ G_{\alpha } (\theta) -  G_{ \alpha ' } (\theta ') }{ G _{ \alpha ' } (\theta ')}
\label{eq:1b}
\end{equation*}

Where, $\theta$ and $\alpha$  represent the RTA and magnetization configuration, namely  parallel (P) or anti-parallel (AP),  respectively.
In this case, the magneto-resistance (MR) is nothing but $ MR (\theta) = \left(\frac{\Delta G}{G}\right)_{\alpha,-\alpha } (\theta )$ where the RTA is fixed and the magnetization direction 
($\alpha$) is changed from P to AP. Similarly, the Twisting resistance is nothing but $ TwR_{\alpha} = \left(\frac{\Delta G}{G}\right)_{\alpha} (\theta, \theta ' )$  where the magnetization 
direction is fixed either to parallel or anti-parallel and the RTA is varied.

\begin{figure*}[htp]
\centering
\includegraphics[clip = true, trim=0cm 0.0cm 0cm 0.0cm, width=1.0\textwidth]{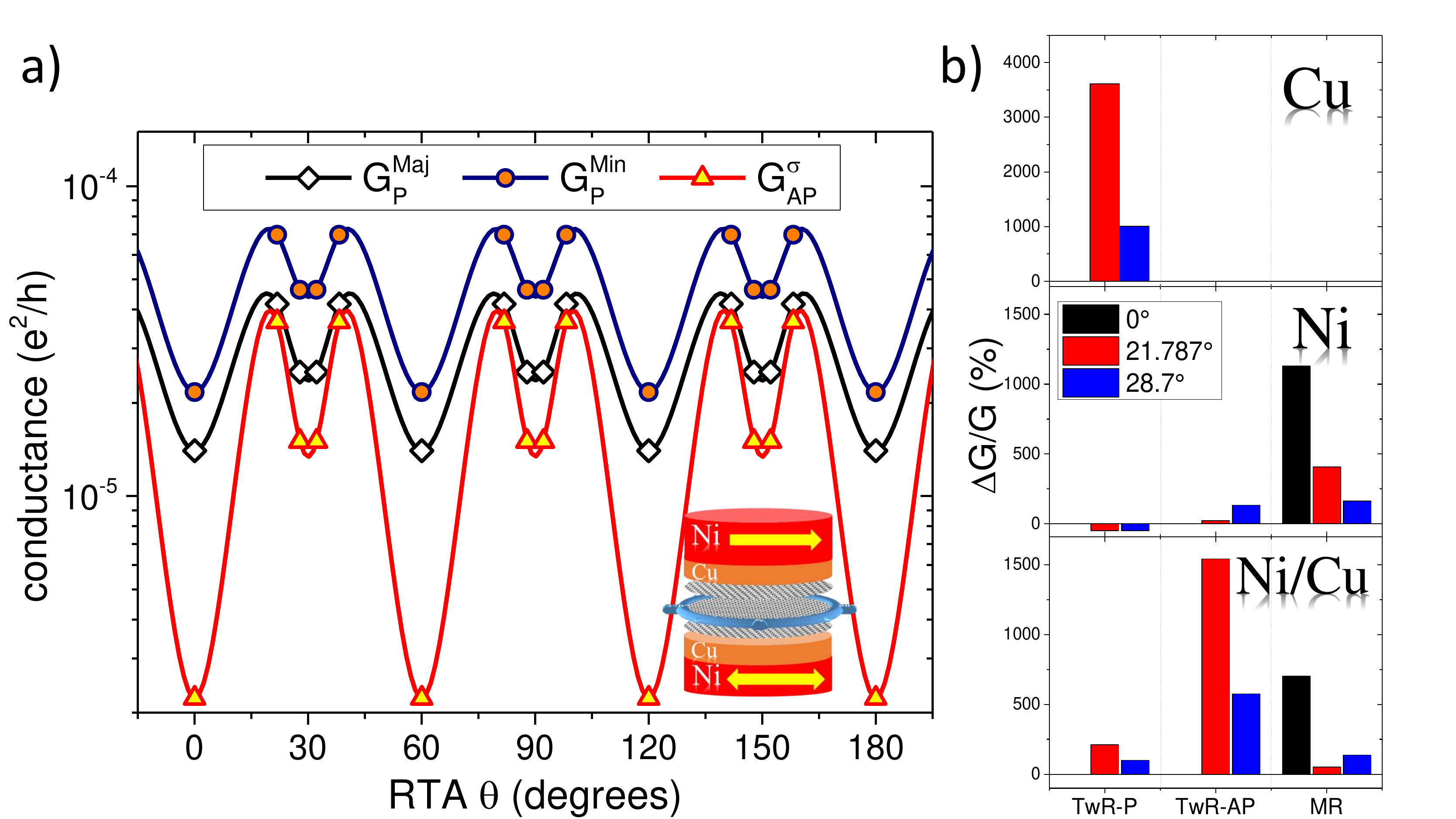}
\caption{ (a) Majority parallel, minority parallel and anti-parallel conductances of Ni/Cu(3ML)/TTLG/Cu(3ML)/Ni MTJs at room temperature as function of the RTA with a schematic of 
the device is shown in the inset. (b) twisting magneto-resistance of the parallel and anti-parallel configurations with different RTA for the different investigated metallic substrate, 
Cu (top), Ni(middle), and Ni/Cu (bottom).}
\label{3}
\end{figure*}

Next, we discuss the non-magnetic case (Cu/TTLG/Cu) by looking at the variation of quantum conductance at Fermi energy and at room temperature as function of RTA ($\theta$) as shown 
Fig.~\ref{1} (d).  The conductance show an increases of almost two order of magnitude for RTA of 21.787$^{\circ}$ compared to the non-twisted case and then slightly decreases for 
$\theta =$ 27.8$^{\circ}$. The estimated TwR are found to be about 3600\% and 1000\% for RTA of 21.787$^{\circ}$ and 28.7$^{\circ}$, respectively. These large values,  comparable 
to the tunneling magento-resistance found in Fe/MgO MTJs, pave the way towards using the mechanical degree of freedom  in storage and logic devices. To understand the origin of 
these large values we plot in Fig.~\ref{1} (e) the Transmission profile across the Brillouin zone for different angles. For the non-twisted case, the Transmission are quite homogeneous 
across the Brillouin zone except at the high symmetry point K where the conductance is minimal. This behavior is expected since the pure Cu have no state present at these states as shown 
in Fig.~\ref{1} (a). As the middle graphene layer is rotated the transmission probability at K-point become maximal and contribute strongly to the conductance ( c.f. Fig.~\ref{1} (e)  middle and bottom panel). 

For the magnetic case, i.e in Ni/TTLG/Ni junctions (Fig.~\ref{2} (a)), the transport properties are influenced by the relative magnetizations configuration in addition to the mechanical degree of freedom arising 
from the TTGL spacer. The conductance in case of majority ($G_{p}^{Maj}$) , minority parallel($G_{P}^{Min}$) and anti-parallel($G_{AP}^{\sigma}$) configuration are shown in Fig.~\ref{2} (b) as function of 
RTA $\theta$. In the anti-parallel case, ($G_{AP}^{\sigma}$ increases with increasing the RTA angle where it reach maximum around 30$^{\circ}$. Interestingly, $G_{p}^{Maj}$ ($G_{P}^{Min}$) show a similar 
(opposite) behavior to that of $G_{AP}^{\sigma}$ where it increases (decreases) as RTA increases from 0$^{\circ}$ to 30$^{\circ}$. Thus, the overall parallel conductance is less effected by the RTA compared 
to the anitparallel case leading to a decrease by one order of magnitude of MR from 1000\% at RTA = 0$^{\circ}$ to 163\% at RTA =27.8$^{\circ}$. Therefore, the spin filtering effect of graphite is sensitive 
to the relative rotation angle of its layer. The magnitude of TwR found in case of Ni is much smaller than that found in Cu substrate. However, the sign of TwR depends on the alignment of magnetizations, 
if it parallel TwR$ < 0$ and if they are anti-parallel TwR is $ > 0$. 
The transmission profile for majority parallel, minority parallel, and anti-parallel configuration at Fermi energy across the Brillouin zone for different angles is plotted in Fig.~\ref{2} (c). Similar to 
the Cu case, the most pronounced changes as function of RTA occurs around the special point in Brillouin zone K. For RTA = 0$^{\circ}$, as expected and shown previously in the literature the transmission 
probability in the parallel configuration is mostly dominated by the spin-down channel (Min-P) around the K point.  For the spin-up channel (Maj-P) the transmission probability across the Brillouin zone is 
quite negligible compared to that of Min-P. This lead to the predicted perfect spin filtering effect in graphene based tunnel junction. For RTA $\not= 0^{\circ} $ the Dirac point of the middle layer is not 
anymore positioned in top of the K point of the Ni. Thus in this case, the transmission of Maj-P and AP will increase while that of Min-P will decreases around the K point as shown in Fig.~\ref{2} (c) left, 
right, and middle panel, respectively.

To enhance the TwR in the Ni/TTLG/Ni case, we propose to introduce a thin Cu film at the interface between Ni and Gr which will increase the TwR without affecting too much the MR. Fig.~\ref{3} (a) 
show the conductance at room temperature as function of RTA for Ni/Cu(3ML)/TTLG/Cu(3ml)/Ni. Overall the insertion of Cu film decreases the conductivity of the system by one order of magnitude, and 
unlike pure Ni case the $G^{Min} _P$ show an increase in conductance value as function of RTA. The latter change of behavior led to increases and changes the sign of TwR in the parallel configuration 
from negative to positive values where TwR-P in this case is about 200\% ( c.f. Fig~\ref{3}). Moreover,  the twisting-resistance in the anti-parallel configuration (TwR-AP) is one order of magnitude 
larger than the case where no Cu inserted and reaches a values of 1542\% and 575\%  for  RTA of 21.78$^\circ$ and 27.8$^\circ$, respectively.  At the same time, the MR  decreases to  more than one order 
of magnitude from 700\% at $\theta =0^\circ$ to 52\% at $\theta = 21.787 ^\circ$. This indicate that the spin-filtering effect of graphene based MTJs are strongly effected by the crystal structure and 
orientation of the spacer. 

These findings are promising for next generation sensor and MTJs based on 2D materials with novel mechanism. One of the challenges in such devices is to control the RTA at the nanoscale in twisted multi-layer 
graphene. Recent experimental work showed the possibility to modify the RTA in-situ with an atomic force microscope tip~\cite{Ribeiro-Palau690}, however the transition speed is slow for plausible applications. 
Other alternative is to use optical nano-tweezers to trap a graphene layer and twisted with respect to substrate, recently graphene trapping is achieved at the microscale using optoelectronic tweezer~\cite{Lim2018}. 
With the fast developing research of nano-optical tweezer by trapping and rotating nanostructures~\cite{Wang2011,Marago2013}, the possibility to control RTA of multi layer graphene at the nanoscale should be within 
the reach in the few next years. Finally, theoretical work predicted that thin topological insulator with broken inversion symmetry might experience a torque or "twisting force"~\cite{Maghrebi2019} as nonequilibrium 
response to temperature difference with the environment. This phenomena could be used to control efficiently the mechanical degree of freedom in graphene based-MTJs.

In conclusion, we demonstrate, Using spin-dependent transport calculations,  the  mechanical control of electric current by twisting the graphene middle layer in a trilayer graphene spacer between two 
metallic leads (magnetic and non-magnetic). A large difference in conductance has been found between two RTA where the TwR could reach more than 1000\% for Cu/TTLG/Cu junctions. For the Ni/TTLG/Ni case, 
we show that the magneto-resistance decreases when the middle layer is twisted with respect to those at the interface, and the TwR depends strongly on the relative magnetization configuration. For the parallel 
configuration, the TwR is about -40\%, while for the anti-parallel configuration it can reach up to 130\% .Furthermore, we report a heterostructure composed of Ni/Cu/TTLG/Cu/Ni where the TwR for parallel and 
anti-parallel configurations can reach up to 200\% and 1600\%, respectively.  Those findings pave the way towards the integration of 2D materials in novel spinmechatronics based devices.

\section{acknowledgments}
The authors would like to acknowledge Prof. Mairbek Chshiev and Dr. Fatima Ibrahim for the fruitful discussion and reviewing the manuscript.
This project has received funding from the European Union Horizon 2020 research and innovation Programme under grant agreement No. 785219 (Graphene Flagship).


\bibliography{twist}

\end{document}